\documentclass[12pt,prb,superscriptaddress]{revtex4}
\usepackage{amssymb}
\usepackage{graphicx}
\usepackage{times}

\begin{document}

\title{Coherent oscillations between classically separable quantum states of
a superconducting loop}
\author{Vladimir E. Manucharyan}
\author{Jens Koch}
\author{Markus~Brink}
\author{Leonid I. Glazman}
\author{Michel H. Devoret}
\affiliation{Departments of Physics and Applied Physics, Yale University, New Haven,
Connecticut USA}
\maketitle


\textbf{Ten years ago, coherent oscillations between two quantum states of a
superconducting circuit differing by the presence or absence of a single
Cooper pair on a metallic island were observed for the first time\cite%
{Nakamura99}. This result immediately stimulated the development of several
other types of superconducting quantum circuits behaving as artificial
\textquotedblleft atoms\textquotedblright \cite%
{Vion02,Martinis02,Chiorescu03,Schrier08,Hoskinson09}, thus bridging
mesoscopic and atomic physics. Interestingly, none of these circuits fully
implements the now almost 30 year old proposal of A. J. Leggett\cite%
{Leggett80} to observe coherent oscillations between two states\textbf{\ }%
differing by the presence or absence of a single fluxon trapped in the 
\textbf{superconducting loop interrupted by a Josephson tunnel junction}.
This phenomenon of reversible quantum tunneling between two classically
separable states, known as Macroscopic Quantum Coherence (MQC), is regarded
crucial for precision tests of whether macroscopic systems such as circuits
fully obey quantum mechanics\cite{Leggett02, TakagiMQTBook}. In this
article, we report the observation of such oscillations with sub-GHz
frequency and quality factor larger than }$\mathbf{500}$\textbf{. We
achieved this result with two innovations. First, our ring has an inductance
four orders of magnitude larger than that considered by Leggett, combined
with a junction in the charging regime\cite{Manucharyan09}, a parameter
choice never addressed in previous experiments\cite{MQCexperiments}. The
higher the inductance and the smaller the capacitance of the small junction,
the smaller the sensitivity of the spectrum of the }\textquotedblleft 
\textbf{atom\textquotedblright\ to variations in the externally applied flux
in the ring. Second, readout is performed with a novel dispersive scheme
which eliminates the electromagnetic relaxation process induced by the
measurement circuit (also known as Purcell effect\cite{Purcell}). Moreover,
the reset of the system to its ground state is naturally built into this
scheme, working even if the transition energy is smaller than that of
temperature fluctuations. As we argue in this article, the MQC transition
could therefore be, contrary to expectations, the basis of a superconducting
qubit of improved coherence and readout fidelity.}

When Anthony Leggett wrote in 1980: \textquotedblleft \lbrack ...] at the
time of writing I\ am inclined to believe that it will turn out to be
impossible in practice (at least in the near future) to see `full-blooded'
coherence between states [of a Radio Frequency (RF)-SQUID] differing by a
full flux quantum...\textquotedblright \cite{Leggett80}, two major
decoherence sources seemed very difficult to circumvent: (i) Damping of the
MQC oscillations by the readout circuitry and (ii) noise in the magnetic
flux threading the loop. The spontaneous transition rate between two states $%
\left\vert 0\right\rangle $ and $\left\vert 1\right\rangle $ of a Josephson
junction, induced by its coupling to the total admittance $Y\left( \omega
\right) $ of the electromagnetic environment between the junction terminals,
is given by $\Gamma _{1\rightarrow 0}=2\left\vert \left\langle 0\left\vert 
\hat{\varphi}\right\vert 1\right\rangle \right\vert ^{2}R_{Q}$\textrm{Re}$%
\left[ Y\left( \omega _{01}\right) \right] \omega _{01}$\cite{Schoelkopf02},
where $R_{Q}=\frac{\hbar }{\left( 2e\right) ^{2}}\simeq 1~\mathrm{k}\Omega $
is the resistance quantum for Cooper pairs, and $\hat{\varphi}$ is the gauge
invariant phase difference operator across the junction. For the two lowest
energy states of a RF-SQUID biased at the half-flux-quantum sweet spot $\Phi
_{\mathrm{ext}}=\Phi _{0}/2$, on which we concentrate in this work, the
matrix element $\left\langle 0\left\vert \hat{\varphi}\right\vert
1\right\rangle $ almost coincides with half the $2\pi $-travel of phase in
the classical limit. For magnetometry measurements of the phase state using
a second SQUID, \textrm{Re}$\left[ Y\left( \omega \right) \right] $ tends to
be of order $1/Z_{\mathrm{vac}}$, where $Z_{\mathrm{vac}}\simeq 377~\Omega $
is the vacuum impedance, and since $R_{Q}>Z_{\mathrm{vac}}$ the coherence is
very short lived\cite{vanderWal00, Friedman00}. Turning now to flux noise
induced decoherence, even at the sweet spot, where first order effects
vanish, a flux fluctuation $\delta \Phi _{\mathrm{ext}}$ will cause a
variation in $\hbar \omega _{01}$ given by $\left( \delta \Phi _{\mathrm{ext}%
}\Phi _{0}/L\right) ^{2}/\left[ \hbar \omega _{01}\left( \Phi _{\mathrm{ext}%
}=\Phi _{0}/2\right) \right] $ where $L$ is the total inductance of the
SQUID loop. Typical values of flux noise and loop inductance in RF-SQUIDs
greatly reduce the coherence of these devices. Nevertheless, coherent
oscillations in the three and four junction flux qubits were finally observed%
\cite{Chiorescu03} and improved to a large extent \cite{Yoshihara06}, but at
the expense of working with non-separable states, i.e. states with largely
overlapping probability distributions, an aspect resulting from the large
tunneling frequencies $\omega _{01}(\Phi _{\mathrm{ext}}=\Phi _{0}/2)$,
typically exceeding $5~\mathrm{GHz}$ in these experiments. In the novel
fluxonium circuit\cite{Manucharyan09}, presented in Fig. 1 (a-c), both
difficulties (i) and (ii) are remedied without sacrifying the key features
of tunneling between separable states and the phase travel of $2\pi $.

\begin{figure}[tbp]
\centering
\includegraphics[width = 0.5\columnwidth]{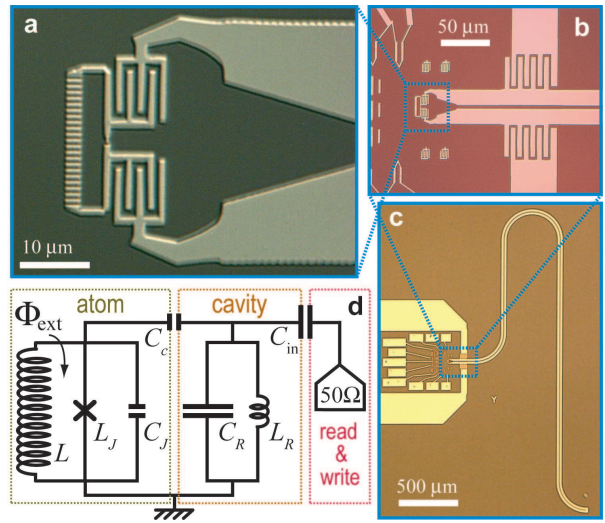}
\caption{\textbf{Fluxonium circuit.} \textbf{a-c} Optical photographs of the
circuit at different scales. Panel \textbf{a} shows a loop consisting of a
small junction shunted by an array of larger area junctions (oriented
vertically) playing the role of an inductance. A pair of interdigitated
capacitances couple the small junction to the end of a \textquotedblleft
parallel wires\textquotedblright\ (i.e. coupled microstrip) transmission
line resonator. Panels \textbf{b} and \textbf{c} show the larger scale
interdigitated capacitors which set the quality factor $400$ of the
resonator through the leakage of signals to the coupled microstrip $%
50~\Omega $ measurement line. Panel \textbf{c} shows the entire snaked
length of the quarter-wavelength resonator whose resonance frequency is $%
8.175~\mathrm{GHz}$. Also shown in this panel are test structures lying in
the gap of the measurement line. Panel \textbf{d} represents the minimal
circuit model of the device, which emulates an \textquotedblleft
atom+cavity\textquotedblright\ system. In this model, the array is
represented as a single ideal inductance $L$ while the distributed resonator
is represented as a parallel combination of resonator inductance $L_{R}$ and
capacitance $C_{R}$, with resonance frequency $\protect\omega %
_{R}=(L_{R}C_{R})^{-1/2}$. The interdigitated capacitors are modelled by
capacitances $C_{c}$ and $C_{\mathrm{in}}$. The coupling between the
\textquotedblleft atom\textquotedblright\ and the \textquotedblleft
cavity\textquotedblright\ is given by the linear interation term $g\frac{%
\hat{Q}}{2e}\left( a+a^{\dag }\right) $ with the coupling constant $g=%
\protect\omega _{R}\frac{C_{c}}{\left( C_{J}+C_{c}\right) }\protect\sqrt{%
\frac{Z_{R}}{R_{Q}}}\simeq 2\protect\pi \times 135~\mathrm{MHz}$, $Z_{R}$
being the oscillator impedance given by $\protect\sqrt{L_{R}/C_{R}}\approx
80 $ $\Omega $. The device is placed in an external magnetic field and the
applied flux threading the loop is $\Phi _{\mathrm{ext}}$, which is close to
a half flux quantum $\Phi _{0}/2$.}
\label{fig: Fig1}
\end{figure}

The equivalent electrical circuit of the fluxonium (Fig. 1d) consists of a
small junction with Josephson inductance $L_{J}$ and capacitance $C_{J}$
shunted by an inductance $L$ provided by a series array of carefully chosen
larger area tunnel junctions, which is approximately $10,000$ times more
inductive than a wire of the same length ($20$ $\mu \mathrm{m}$). In order
to read the states of such circuit, an operation described in detail below,
the junction is coupled via a capacitance $C_{c}$ to a $L_{R}C_{R}$ harmonic
oscillator implemented with a quarter-wave transmission line resonator (Fig.
1c). In the resulting \textquotedblleft atom+cavity\textquotedblright\ system%
\cite{Raimond01,Wallraff04}, the dynamical variables of the
\textquotedblleft atom\textquotedblright\ are the flux $\Phi $ across the
inductance $L$\cite{DevoretHouchesFluctuations}, and its canonically
conjugate variable $Q$, which coincides here with the charge on the junction
capacitance. Note that in contrast with the position and momentum of an
electron in an ordinary atom, here the pair of variables $\Phi $ and $Q$
describe the collective motion of a superconducting condensate around an
entire circuit loop. Quantum-mechanically, these variables must be treated
as operators satisfying $[\hat{\Phi},\hat{Q}]=i\hbar $ and the cavity mode
is described by annihilation and creation operators $\hat{a}$ and $\hat{a}%
^{\dag }$ with $\left[ \hat{a},\hat{a}^{\dag }\right] =1$. The
\textquotedblleft atom\textquotedblright\ is characterized by the three
energies measured in a previous experiment on the same device but near zero
external field: the Josephson energy $E_{J}/h=\frac{1}{h}(\Phi _{0}/2\pi
)^{2}/L_{J}=8.9~\mathrm{GHz}$\textrm{, }the Coulomb charging energy $E_{C}/h=%
\frac{1}{h}\frac{1}{2}e^{2}/\left( C_{J}+C_{c}\right) =2.4~\mathrm{GHz}$ and
the inductive energy $E_{L}/h=$ $\frac{1}{h}(\Phi _{0}/2\pi )^{2}/L=0.52~%
\mathrm{GHz}$. A conservative estimate of the uncertainty in the values
given here is $2\%$, $8\%$ and $2\%$, respectively. These three energies
place our circuit in a so far unexplored niche of Josephson devices where $%
E_{J}/E_{C}$ is of order unity while $E_{J}/E_{L}$ is much larger than unity
(see Fig. 2a). In this niche, like in that of SQUID-like devices, charge
noise is suppressed, but remarkably, the sensitivity of the spectrum to flux
noise is also greatly reduced\cite{Koch09}.

\begin{figure}[tbp]
\centering
\includegraphics[width = 0.5\columnwidth]{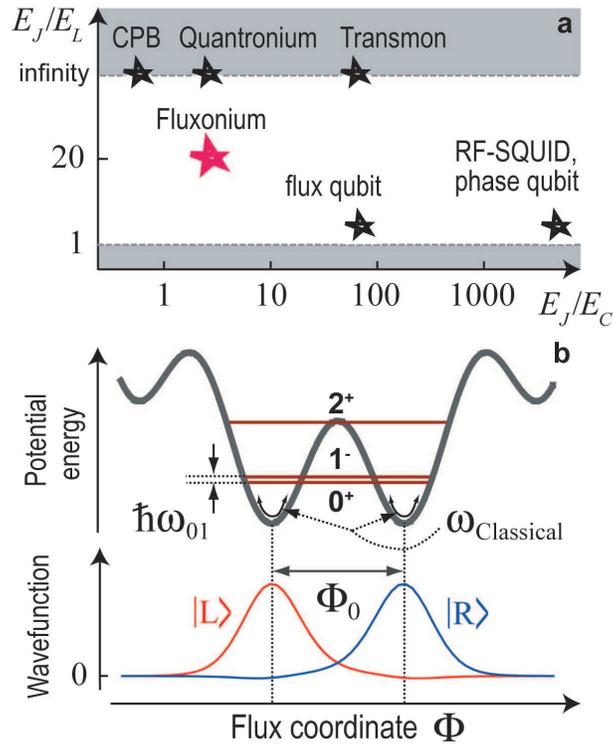}
\caption{\textbf{Specificity of fluxonium.} Panel \textbf{a }compares the
parameters of the fluxonium artificial atom to that of other superconducting
artificial atoms (qubits). For the Cooper pair box, quantronium and
transmon, the effective inductance $L$ can be considered infinite. The
RF-SQUID and the phase qubits are represented by one star, because their
parameters nearly coincide. For the flux qubit, the role of the shunting
inductance can be thought of as that provided by the Josephson inductance of
two or three larger junctions in series. Panel \textbf{b} shows the
potential landscape seen by the flux coordinate $\Phi $ of the
\textquotedblleft atom\textquotedblright\ of Fig. 1 for $\Phi _{\mathrm{ext}%
}=\Phi _{0}/2$. The two lowest minima, separated by a barrier and spanning a
flux quantum, define two classically separable states. A necessary condition
for their existance is $E_{J}/E_{L}>1$. Such states, in the classical limit
of $h\rightarrow 0$ correspond to vibrations near the bottoms of the minima
with the characteristic frequency $\protect\omega _{\mathrm{Classical}}$.
The splitting between the two lowest quantum levels ($0^{+}$ and $1^{-}$) is
due to reversible quantum tunneling through the barrier and define the
\textquotedblleft macroscopic quantum tunneling\textquotedblright\ frequency 
$\protect\omega _{\mathrm{01}}$. The wavefunctions for the levels $0^{+}$
and $1^{-}$ are symmetric and antisymmetric superpositions of the $%
\left\vert L\right\rangle $ (left) and $\left\vert R\right\rangle $ (right)
wavefunctions plotted in the lower half of the panel. The latter two
wavefunctions barely overlap under the barrier, indicating that tunneling is
strongly suppressed. The third level ($2^{+}$) lies above the barrier and
the transition 1-2, whose frequency is of order $\protect\omega _{\mathrm{%
Classical}}$ and about $30$ times larger than that of transition 0-1, is
used for reading out which of levels $0^{+}$ and $1^{-}$ is occupied.}
\label{fig: Fig2}
\end{figure}

Before describing our measurement, let us discuss quantitatively the issue
of state separability. When the loop is biased by an external flux close to
the half-flux-quantum $\Phi _{0}/2$, the two lowest states $\left\vert
0\right\rangle $ and $\left\vert 1\right\rangle $ of the fluxonium are the
symmetric and antisymmetric combination of the states $\left\vert
L\right\rangle $ and $\left\vert R\right\rangle $ described by two real
wavefunctions $\left\langle \Phi |L\right\rangle $ and $\left\langle \Phi
|R\right\rangle $ plotted in Fig. 2b. They are localized at $\Phi =-\Phi
_{0}/2$ and $\Phi =+\Phi _{0}/2$, which are the location of the left and
right minima of the double well potential seen by the flux coordinate $\Phi $%
, respectively. We consider the two states $\left\vert L\right\rangle $ and $%
\left\vert R\right\rangle $ classically separable because when in state $%
\left\vert L(R)\right\rangle $, the probability to find the system in the
right (left) well is much less than unity, or in more quantitative terms $%
s=\left\vert \left\langle 0\left\vert \Phi \right\vert 1\right\rangle
\right\vert ^{2}/\sqrt{\sigma _{L}\sigma _{R}}\gg 1$ where $\sigma _{\Psi
}^{2}=\left\langle \Psi \left\vert \Phi ^{2}\right\vert \Psi \right\rangle
-\left\langle \Psi \left\vert \Phi \right\vert \Psi \right\rangle ^{2}$; $%
s=8.5$ in the present experiment. Time domain coherent oscillations between
these two states of a superconducting loop qualify as the MQC phenomenon. An
equivalent way to evaluate the degree of separability is to compare the MQC
oscillation frequency $\omega _{01}$ with the frequency of classical
oscillations in either well, given approximately by $\omega _{\mathrm{%
Classical}}=\sqrt{8E_{J}E_{C}}/h$. In the present experiment the two
frequencies $\omega _{01}/2\pi $ and $\omega _{\mathrm{Classical}}/2\pi $
are predicted, with better than $10\%$ accuracy, to be given by $353~\mathrm{%
MHz}$ and $13.5~\mathrm{GHz}$, respectively. Let us note that the center of
mass motion of an ammonia molecule undergoing coherent tunneling (inversion
transition) is equally qualified as an oscillation between two classically
separable states in the sense we have given above\cite{AmmoniaMolecules}
(see also a closely related discussion of Hund's paradox\cite%
{ChiralMolecules}).

How does one go about measuring the MQC oscillations? Directly measuring the
flux generated by the MQC states with a SQUID is not an option here, despite
the maximal swing in $\Phi $: the mutual inductance of the fluxonium loop to
any other superconducting loop in the vicinity - typically a $\mathrm{pH}$
per $\mathrm{\mu m}$ of wire - would approximately be $10,000$ times smaller
than the fluxonium loop inductance $L=300~\mathrm{nH}$. Moreover, the
amplitude of the current generated in the loop by the MQC oscillations is
only of order $1~\mathrm{nA}$\textrm{, }so that in the end only a flux
oscillation of $10^{-5}\Phi _{0}$ would be measured by a readout SQUID. We
have circumvented these problems by a capacitive measurement scheme that
exploits the presence of the second excited state, which lies slightly above
the barrier of the double well (see Fig. 2b). The resonator frequency $%
\omega _{R}$ is chosen to be close to the 1-2 transition frequency of the
atom, and their interaction provides a way to monitor dispersively the
atomic 0-1 transition, as we show below. This new type of superconducting
qubit readout shares common features with recent optical QND preparation and
readout of spin-squeezed hyperfine clock states of Rb and Cs atoms\cite%
{Polzik,Vuletic}. Remarkably, the large separation of the cavity and qubit
transition frequencies, $\omega _{01}\ll \omega _{12},\omega _{R}$, forbids
in our scheme the spontaneous emission of a photon from the qubit first
excited state into the cavity (Purcell effect). In other words, \textrm{Re}$%
\left[ Y\left( \omega _{01}\right) \right] $ is minimized without
jeopardizing the readout fidelity.

\begin{figure}[tbp]
\centering
\includegraphics[width = \columnwidth]{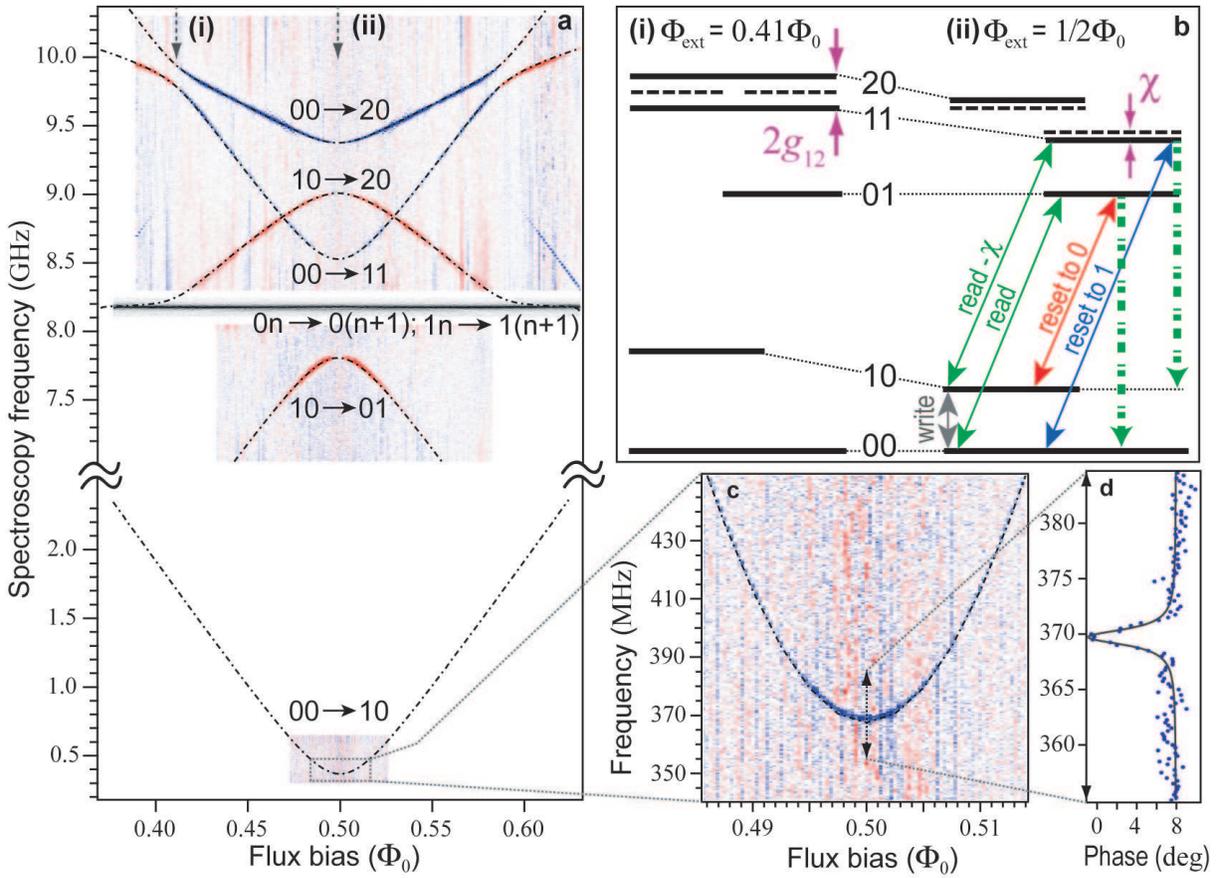}
\caption{\textbf{Two-tone spectroscopy.} Panel \textbf{a} shows the phase of
the reflected readout tone as a function of the spectroscopy tone frequency
and external flux. The color scale encodes the value of the phase, with zero
corresponding to the mauve background, blue to negative values (dips), and
red to positive values (peaks). The gray region around $8.2~\mathrm{GHz}$
shows the reflected phase of a single tone, swept close to the resonator
bare frequency. Theoretical predictions are shown in dot-dash lines.
Transition assignments are indicated with a two-digit code, the first digit
corresponding to the atom state and the second digit corresponding to the
cavity photon number, as clarified in panel \textbf{b}, which shows the
various functions of the irradiations used in the experiment. In that latter
panel, the combined atom-cavity levels are shown for two special values of
flux bias: the 11-20 degeneracy point (i) and the half flux quantum sweet
spot (ii). Dashed lines indicate levels 11 and 20 in the absence of
atom-cavity interaction. Panels \textbf{c} and \textbf{d} expand the sub-GHz
MQC transition (00-10) observed in the vicinity of the sweet spot.}
\label{fig: Fig3}
\end{figure}

We now turn to the two-tone spectroscopy results presented in Fig. 3a. They
were obtained by applying a fixed frequency readout tone at the cavity
frequency, and measuring the phase of the reflected readout signal as a
function of the frequency of a second tone exciting the atomic transitions%
\cite{Schuster05}. The flux dependence of all transition frequencies of the
combined \textquotedblleft atom+cavity\textquotedblright\ system agrees
perfectly with theoretical predictions based on five adjustable parameters.
The states are labeled with two numbers, the first referring to the atomic
excitation and the second referring to the cavity excitation. We observe the
atomic transitions 00-10, 10-20 and 00-20, as well as the red and blue
sidebands involving both atom and cavity (10-01 and 00-11). In the same
panel, a straight grey fuzzy line correspond to the measurement of the
cavity response, obtained by single tone spectroscopy. This data notably
displays the novel strong interaction regime of coupling between states 02
and 11, which manifests itself by the avoidance of the transitions 00-11 and
00-20 at the frequency of about $10~\mathrm{GHz}$. The minimum splitting $%
2g_{12}$ equals $130~\mathrm{MHz}$. At the same time, the MQC transition
00-10 located far down is surprizingly well resolved, with a power-broadened
linewidth of about $3$~$\mathrm{MHz}$,\textrm{\ }as shown in the fine-scale
spectroscopy data of the panels c) and d). The transition frequency passes
through a minimum at half-flux quantum and the absolute measured value ($%
368.9\pm 0.3)~\mathrm{MHz}$ of this lowest frequency through a Lorentzian
fit is in agreement with theoretical predictions, within error bars.

To explain how the MQC transition, so highly detuned from the cavity, could
be observed at all, we examine the minimal set of five levels represented in
Fig. 3b for the coupled \textquotedblleft atom+cavity\textquotedblright\
system. Two particular values of flux have been chosen, corresponding to
arrows (i) and (ii) in panel a. At the point of maximum coupling (i), the
states 20 and 11 have completely hybridized, while at the half-flux-quantum
sweet spot (ii), where the hybridization is greatly reduced, they still
repel significantly, producing a shift $\chi =\omega _{00\longrightarrow
01}- $ $\omega _{10\longrightarrow 11}$ of the cavity frequency conditioned
by the excitation of the atom in its 1 state. To second order in $g$, $\chi
=\left( g_{12}\right) ^{2}/\left( \omega _{12}-\omega _{R}\right) \simeq 10~%
\mathrm{MHz}$. Thus, it is possible to read out the atom state by
irradiating the system at the cavity frequency. In addition, by irradiating
the red and blue sidebands, we can reset the qubit to either state 0 and 1.
For instance, in the case of red sideband irradiation, we are transferring a
quantum of excitation from the qubit to the resonator, which in turn emits
its energy into the $50~\mathrm{\Omega }$ input impedance of the readout
amplifier. This important built-in reset feature suppresses the usual qubit
requirement $k_{B}T\ll \hbar \omega _{01}$. Both the readout and reset
manipulations are based on the fact that the relaxation rate of the cavity
is much faster than the relaxation rate of the excited state involved in the
MQC transition, i.e. $\kappa \gg \Gamma _{01}$.

Measurements in the time domain are performed by applying the protocol
described in Fig. 3b(ii). A reset pulse is first applied to initialize the
MQC doublet in either ground or excited state, and then a Rabi drive pulse
is applied to write a given superposition state. The results are shown in
Fig. 4a. When initialized in what is supposed to be the ground state, the
system displays Rabi oscillations with a contrast increased relative to the
thermal equilibrium value by a factor of 2, consistent with the estimated
temperature of the sample. A reversal of the phase of Rabi oscillations is
clearly observed when we now reset the system to what is supposed to be the
1 state. However, the contrast of the oscillations is found to be weaker
than for the reset to 0, a discrepancy which we attribute to the fact that
reset requires such power that spurious transitions are likely to occur
since sideband transitions is nearly forbidden at our working point, which
was about $0.05\%$ away from the half-flux quantum symmetry point.

\begin{figure}[tbp]
\centering
\includegraphics[width = \columnwidth]{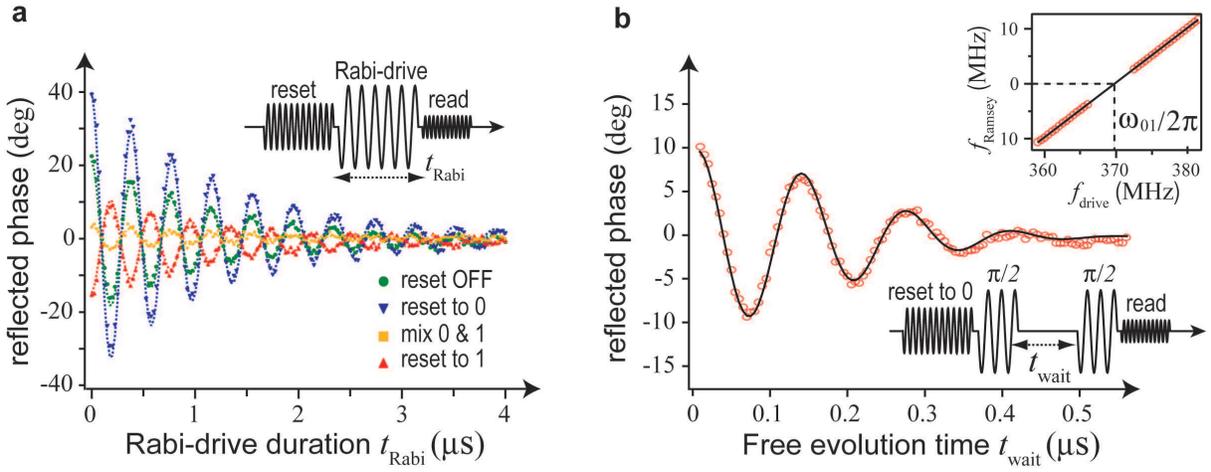}
\caption{\textbf{Observation of MQC in the time domain.} Panel \textbf{a}
shows Rabi oscillations between ground (0) and first excited state (1) of
the qubit, measured for different qubit initialization protocols. Green
dots: wait for thermal equibilibrium no reset. Blue/Red triangle: reset to
0/1 applied (10-01/00-11 transition in Fig. 3b). Yellow squares: reset-to-1
pulse of weaker amplitude applied to nearly maximally mix states 0 and 1.
Panel \textbf{b} shows Ramsey fringe after initialization to ground state,
with inset giving Ramsey beating frequency as a function of qubit drive
frequency. The fringes correspond to stroboscopy of MQC oscillations between
states $\left\vert L\right\rangle $ and $\left\vert R\right\rangle $.}
\label{fig: Fig4}
\end{figure}

Finally, the coherence of the MQC oscillation is measured using the Ramsey
fringe protocol, as shown in Fig. 4b. After a reset pulse to the ground
state, two $\pi /2$ pulses separated by a free-evolution waiting time are
applied to the sample. The protocol ends with a final measurement pulse. We
have taken this data for different drive frequencies as shown in the inset,
confirming that the beating frequency $f_{\mathrm{Ramsey}}$ is correctly
related to the drive frequency of our $\pi /2$ pulses. The fringes have a
Gaussian decay envelope with a characteristic decay time of $T_{\mathrm{%
Ramsey}}=250~\mathrm{ns}$, corresponding to a coherence quality factor $%
Q_{MQC}=T_{\mathrm{Ramsey}}\omega _{01}=580$. At the times corresponding to
the extrema of the fringes, the systems is passing through the classically
separable states $\left\vert L\right\rangle $ and $\left\vert R\right\rangle 
$. The decay time of the fringes is much shorter than the relaxation time $%
T_{1}>$ $5~\mathrm{\mu s}$, obtained in a separate experiment. At the time
of this writing, we do not yet have an explanation for this short coherence
time: At the transition frequency minimum, simple predictions based on the $%
10^{-6}\Phi _{0}/\left( \mathrm{Hz}\right) ^{1/2}@1~\mathrm{Hz}$ $1/f$ flux
noise limitations observed at larger flux bias and treated here to second
order, as well as estimates based on critical current $1/f$ noise
fluctuations $10^{-6}I_{0}/\left( \mathrm{Hz}\right) ^{1/2}@1~\mathrm{Hz}$ 
\cite{VanHarlingen04} of the small junction give $T_{\mathrm{Ramsey}}>10~%
\mathrm{ms}$ and $T_{\mathrm{Ramsey}}>100~\mathrm{\mu s}$, respectively.
Another concern would be fluctuations in the value of the inductance of the
array; however $d\omega _{01}/dL$ is minimal at half-flux-quantum whereas
measured coherence improves away from this point. Further experiments are
clearly needed to explore other hypotheses like out-of-equilibrium
quasiparticles or quantum phase slips in the junction array\cite{Matveev02}
resulting in decoherence from charge noise. Both of these mechanisms could
be remedied: quasiparticle traps could be added and phase slips could be
suppressed by a slight increase in the array junction size. Furthermore by
using the device in clusters\cite{Gladchenko09}, a system with a
topologically protected doublet ground state could in principle be
implemented, in order to suppress more completely the effect of decoherence%
\cite{Ioffe02}.

In conclusion, we have reported the first observation, in the time domain,
of the reversible tunneling of the macroscopic superconducting phase
difference between two wells of the Josephson potential. In contrast with
experiments on RF-SQUIDs and flux qubits, the total phase travel is $2\pi $
while the standard deviation of the probability distribution at the
oscillation extrema is much less than $\pi $. The oscillation coherence
factor, although surprisingly good in view of initial expectations, is much
lower than what an analysis based on typical levels of noise encountered in
this type of superconducting device predicts. Therefore, the coherence is
likely to improve and this new device would become a very promising qubit
that, while easy to read and to reset, escapes from the limitations of the
Purcell effect: because the frequency of this qubit is one order of
magnitude smaller than that of other superconducting qubits, the control of
noise and dissipation in a reduced frequency range might be easier to
realize, while the time to perform a two-qubit gate would not be slowed
down. Even smaller frequencies are worth exploring, since there is no
limitation on the possibility of artificially resetting the system in its
ground state. These lower frequencies would allow a coupling to
nanomechanical systems \cite{LaHaye09} which themselves are interfacable to
light beams for the transport of quantum information. On a more fundamental
level, our experiment on a fluxonium version of the MQC system confirms its
status as a testing ground for eventual limitations of quantum mechanics,
since it is one of the very few that lends itself to a controlled, absolute
measurement of anomalies in the vanishing reversible tunneling rate between
two classically separable macroscopic states.

We thank R. Vijay, M. Metcalfe, C. Rigetti, D. Schuster, L. DiCarlo, J.
Chow, L. Bishop, L. Frunzio, R. Schoelkopf and S. Girvin for useful
discussions. Assistance of Nick Masluk with the experiment is greatfully
acknowledged. This research was supported by the NSF under grants
DMR-0906498, DMR-032-5580, the NSA through ARO Grant No. W911NF-05-01-0365,
the Keck foundation, and Agence Nationale pour la Recherche under grant
ANR07-CEXC-003. M.H.D. acknowledges partial support from College de France.

\end{document}